\documentclass[twocolumn,aps,pre,10pt,superscriptaddress,notitlepage,floatfix]{revtex4-1}
\usepackage{graphicx,amsmath}
\usepackage{color}
\usepackage{siunitx}
\usepackage{natbib}
\usepackage[version=4]{mhchem}
\usepackage{natbib}
\usepackage{soul}
\usepackage[hidelinks]{hyperref}
\usepackage{balance}

\begin{document}
\title{Missing links as a source of seemingly variable constants in complex reaction networks}
\author{Zachary G. Nicolaou}
\affiliation{Department of Physics and Astronomy, Northwestern University, Evanston, Illinois 60208, USA}
\author{Adilson E. Motter}
\affiliation{Department of Physics and Astronomy, Northwestern University, Evanston, Illinois 60208, USA}
\affiliation{Northwestern Institute on Complex Systems, Northwestern University, Evanston, Illinois 60208, USA}

\begin{abstract}
A major challenge in network science is to determine parameters governing complex network dynamics from experimental observations and theoretical models.  In complex  chemical reaction networks, for example, such as those describing  processes in internal combustion engines and power generators,  rate constant estimates  vary significantly across studies despite substantial experimental efforts.  Here, we examine the possibility that variability in measured constants can be largely attributed to the impact of missing network information on parameter estimation.  Through the numerical simulation of measurements in incomplete chemical reaction networks, we show that unaccountability of network links presumed unimportant (with local sensitivity amounting to less than two percent of that of a measured link)  can create apparent rate constant variations as large as one order of magnitude even if no experimental errors are present in the data.  Furthermore, the correlation coefficient between the logarithmic deviation of the rate constant estimate and the cumulative relative sensitivity of  the neglected reactions was less than $0.5$ in all cases. Thus, for dynamical processes on complex networks, iteratively expanding a model by determining new parameters from data collected under specific conditions is unlikely to produce reliable results.  \\

\noindent DOI: \href{https://doi.org/10.1103/PhysRevResearch.2.043135}{10.1103/PhysRevResearch.2.043135}
\end{abstract}

\maketitle

\setlength{\belowcaptionskip}{-10pt}
\setlength{\baselineskip}{10.7pt}

\section{Introduction}
In the study of real complex network systems, information about the system components is often incomplete, unreliable, or unknown. 
Recently proposed methods to infer the likelihood of links in a network take advantage of node similarities or network structures \cite{2008_Clauset, 2009_Guilerma, 2010_Lu, 2013_Ravasz,Yu_2018,2019_Pech}.
Combined with statistical inference, such methods have demonstrated remarkable success in a variety of real-world networks with heterogeneous errors \cite{2018_Peixoto}.
Other approaches include observing links present in reconstructed networks after perturbations  \cite{2015_Lu},  applying the inverse problem to generative models for networks with predicted hyperbolic structures \cite{2012_Papadopoulos}, and inferring missing nodes from community detection algorithms \cite{2016_Hric}.  
Statistical measures of correlation and causation have also been used to infer links from observations \cite{2012_Wu,2015_Liao,lunsmann2017transition,casadiego2017model}.

In addition to the topology of their connections, links and nodes often carry parameters that encode their properties.  In models of physical networks, such parameters must be determined through measurements of observables, which may be costly and limited.  In some cases, while the topology of the network is known in principle, parameters such as link weights can only be estimated for a relatively small proportion of the network.  Nevertheless, if the network is derived from a knowledge-based model (i.e., it is believed to include all relevant details and is physically well motivated), it is usually assumed that individual parameters can be measured accurately through targeted experiments. 

The key problem in such cases is not to determine which nodes and links are present, but rather,  which of them are most important to include in the model.  ``Weak'' nodes and links may be systematically neglected to reduce the number of parameters and derive minimal models, but it is unclear how these missing elements may affect the estimated parameters, and hence the predicted dynamics, of the parts of the network that are retained. That is, it remains largely unknown what impact missing and deliberately omitted structural information may have  on the relevant dynamical processes predicted or described by a network model.

\begin{figure}[b]
\includegraphics[width=\columnwidth]{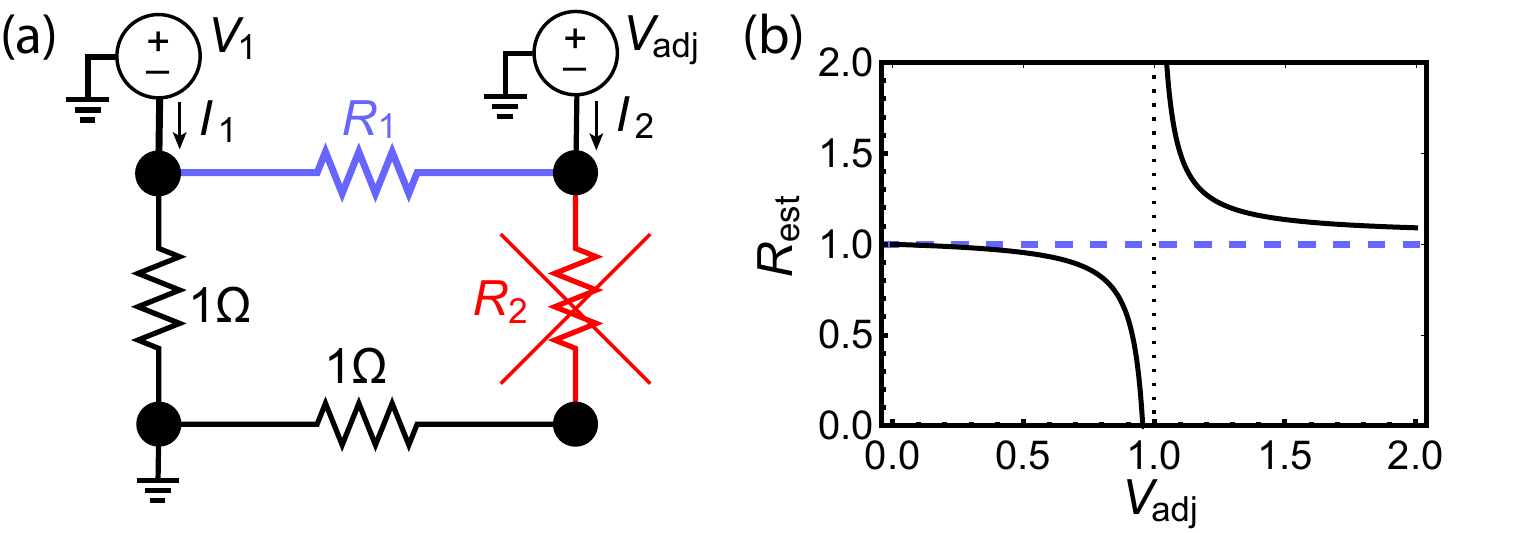}
\caption{Impact of missing information in a simple electric network. (a)~Network of four resistors, where $V_1=1$V is fixed and $V_{\mathrm{adj}}$ differs in different experiments. The resistance of the blue resistor ($R_1=1\Omega$) is unknown to the experimenter and is determined through measurements of the currents $I_{1,2}$ while the red resistor ($R_2=10\Omega$) is missing in the model.  (b)~Estimate of resistor value vs adjustable voltage, as given by the least squares fit. The blue dashed line shows the actual value, and the black dotted line shows where the currents become insensitive to the estimate.
\label{fig1}}
\end{figure}

Here, in contrast to previous studies focused on solely network structure, we explore the impact of network incompleteness on the  parameters governing the system dynamics. To motivate this work, consider the simple electric network of resistors shown in Fig.~\ref{fig1}(a). Suppose an experimenter attempts to fit an incomplete model of the system (in which the red resistor is absent) to current measurements in order to determine the resistance $R_1$ of the blue resistor. Since the missing resistor is relatively large, little current passes through it.  From the network perspective, the links in the network should be given a weight proportional to their admittance (inverse resistance) and, intuitively, a model that neglects small weighted links would still be expected to be good at predicting the dynamics.  The absence of the red resistor results in a systematic error due to the modeling approximation, and we assume for simplicity that the measurements are perfect and no additional errors are present. 

The currents drawn from the voltage sources are measured, and the resistance $R_1$ in the model is adjusted until the model's predicted currents minimize the deviations from the measured currents, resulting in a least squares estimate (see Appendix \ref{acircuit} for details). Figure \ref{fig1}(b) shows how this estimated resistance $R_{\mathrm{est}}$ depends on the voltage value $V_{\mathrm{adj}}$ that the experiment is conducted under. Crucially, two experimenters who operate with different values of $V_{\mathrm{adj}}$ can obtain substantially different estimates for $R_{\mathrm{est}}$ and may falsely conclude that the resistance changes with the experimental conditions. The apparent variability of a constant in this example is actually quite easy to understand. Essentially no current passes through the blue resistor when $V_{\mathrm{adj}}\approx V_1$, so the measured currents are not sensitive to changes in the value of $R_1$ (as quantified by local sensitivity in Appendix \ref{acircuit}). Thus, the error introduced by the incomplete  model is amplified around the dotted line in Fig.~\ref{fig1}(b), and therefore the measure of link importance based on the admittance (link weight) does not adequately reflect the impact of link removal on parameter estimation.

While the errors induced by the estimation procedure in Fig.~\ref{fig1} are easy to understand in terms of local sensitivities, subtler challenges arise in more complex networks, especially when the network dynamics are nonstationary. In the remainder of the paper, we show that for complex chemical reaction networks in particular, missing information about the underlying network can lead to significant variations in rate constants estimated under different conditions even when the local sensitivities for the neglected links are small compared to those for the links being measured.  
Given that network models of chemical reactions will necessarily be incomplete (since additional short-lived radical species and rare reactions could always be added to a given model), this variability poses a serious challenge to the efforts to reliably determine rate constant through experiments.

To illustrate this challenge, we focus on important chemical reaction networks involved in natural gas combustion in Section \ref{combustion} and ethylene pyrolysis in Section \ref{pyrolysis}, which have been carefully developed for decades, primarily through experiments employing shock tubes and laminar flame speed measurements \cite{2003_Simme,2004_Li_Frederick,2004_Conaire_Charles,2005_Baulch,2012_Ranzi}. We compare the results for these processes in Section \ref{comparison}, and find in particular that local sensitivity analysis, which is a commonly used methods for determining reaction importance \cite{1980_Dougherty_Herschel,1990_Turanyi,2013_Tomlin}, correlates weakly with the reactions impact on rate constant measurements in all cases.  We conclude in Section \ref{discussion} with a discussion of implications and directions for future research.

\begin{figure*}
\includegraphics[width=2\columnwidth]{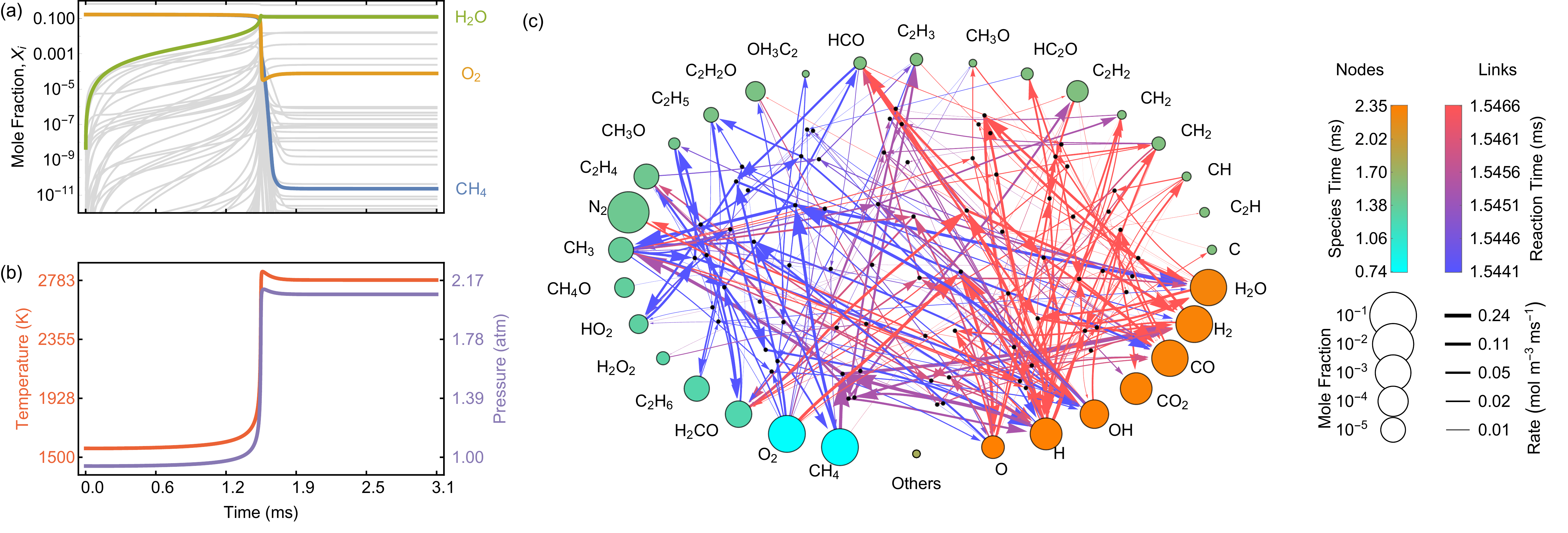}
\caption{{Natural gas ignition in simulations. (a) Mole fractions vs time, with water, methane, and oxygen shown in green, blue, and orange, respectively, and all other species shown in gray. (b) Temperature (red line) and pressure (blue line) vs time, where the sharp rises at $1.5\si{ms}$ corresponds to ignition.(c) Network of chemical species in the natural gas network, where colored nodes show species, black nodes show prominent reactions, and links show reaction flux, with the weight showing the time-integrated reaction rate and the node size showing the time-integrated concentration, both on a logarithmic scale. The node and link color indicate the time that the species and reactions are most prominent, as measured by the time of the maximum reaction flux and the time at which the integrated species  concentration reaches half its ultimate value.  Nodes with identical chemical formulae are isomers, and species with no links weighted above a threshold ($\ce{C3H8}$, $\ce{CH3CHO}$, $\ce{C3H7}$, $\ce{HCCOH}$, $\ce{HCNN}$, $\ce{H2CN}$, $\ce{AR}$, $\ce{HCNO}$, $\ce{HOCN}$, $\ce{HCN}$, $\ce{CN}$, $\ce{HNCO}$, $\ce{NCO}$, $\ce{NH3}$, $\ce{NH2}$, $\ce{NH}$, $\ce{N2O}$, $\ce{N}$, $\ce{HNO}$, $\ce{NO2}$, $\ce{NNH}$, $\ce{NO}$) are shown together in the node labeled ``Others'' with their mean size (and color).  An animation of the temporal evolution of this network is available in the Supplemental Material \cite{SM}.  \label{fig2} }}
\end{figure*}
\section{Natural gas combustion}
\label{combustion}
We first consider the impact of unknown network reactions in the case of natural gas combustion.  
We model the chemical process with a coupled set of kinetic equations, which corresponds to a network of chemical species, elementary reactions, and physical parameters, including the rate constants that relate species concentrations to the rate of reactions (see Appendix~\ref{akinetic}). 
To emulate the effects of missing information in models of real processes, we take a specific network (which we regard as the complete network) as our ``ground truth'' from which measurement values (regarded to be exact) are generated using simulations. 
We employ the GRI 3.0 network \cite{1999_Smith} as our ``complete'' natural gas network, which is then simulated using the open source software Cantera to integrate a continuously stirred reactor with an accurate time stepping method for stiff ordinary differential equations \cite{cantera,SM2}. 
The GRI 3.0 network is a specification of $53$ chemical species and $325$ elementary reactions and rate constants, which have been curated  carefully from the body of experiments in the literature. Despite extensive curation, this network is not reliable for general purposes across different conditions when the parameters are fixed, which our analysis will suggest to be due to incompleteness of the network model itself. But for the purpose of this analysis, we regard this network as complete and evaluate our hypothesis by  quantifying the impact of further removing information from the network. Specifically, starting with this network, we  produce plausible models for the process by removing some reactions from the complete network and use the resulting ``incomplete'' networks to determine parameter values from measurements.

Figure \ref{fig2}(a) shows the evolution of the chemical compositions in a mixture of natural gas and oxygen.  The  reactants combine explosively as the reaction proceeds and ultimate produce water and a variety of other products. During this process, the temperature and pressure of the gas increase sharply as the fuel ignites, as shown in Fig.~\ref{fig2}(b).  This ignition is the result of a chain reaction process taking part in the complex network of species which are connected by elementary reactions.  The network structure of this process can be represented as a hypergraph, where nodes are chemical species and directed hyperedges are reactions, with weights proportional to the reaction flux.  This hypergraph can be visualized by its bipartite incidence graph, where the hyperedges are replaced with reaction nodes, as shown in Fig.~\ref{fig2}(c).

A traditionally employed measure of reaction importance in chemical reaction networks is the local sensitivity, which quantifies changes in  the concentration with changes in the rate constant (see Appendix \ref{akinetic}).  
To assess the reliability of this metric, we generate new incomplete networks from the complete network by randomly removing reactions that are deemed unimportant while retaining reactions presumed to be important, where reaction importance is assessed on the basis of local sensitivities. Rate constants in these incomplete networks are then fit to simulated measurements from the complete network to assess the impact of network incompleteness on parameter estimation.

For our measurement simulations, we consider two reactions whose rate constants are to be determined,
\begin{align}
\ce{H + O_2 <=> O + OH},  \label{eqn1} \tag{I} \\
\ce{CH_4 + H <=> CH_3 + H_2},  \label{eqn2} \tag{II}
\end{align}
taking all other rate constants at the fixed values given in the natural gas network.  Equation \eqref{eqn1} is a chain branching reaction, which helps to populate the radical pool that leads to ignition.  Equation \eqref{eqn2} is a propagation reaction, which helps to diversify this pool of radicals. 

\begin{figure}[b]
\includegraphics[width=\columnwidth]{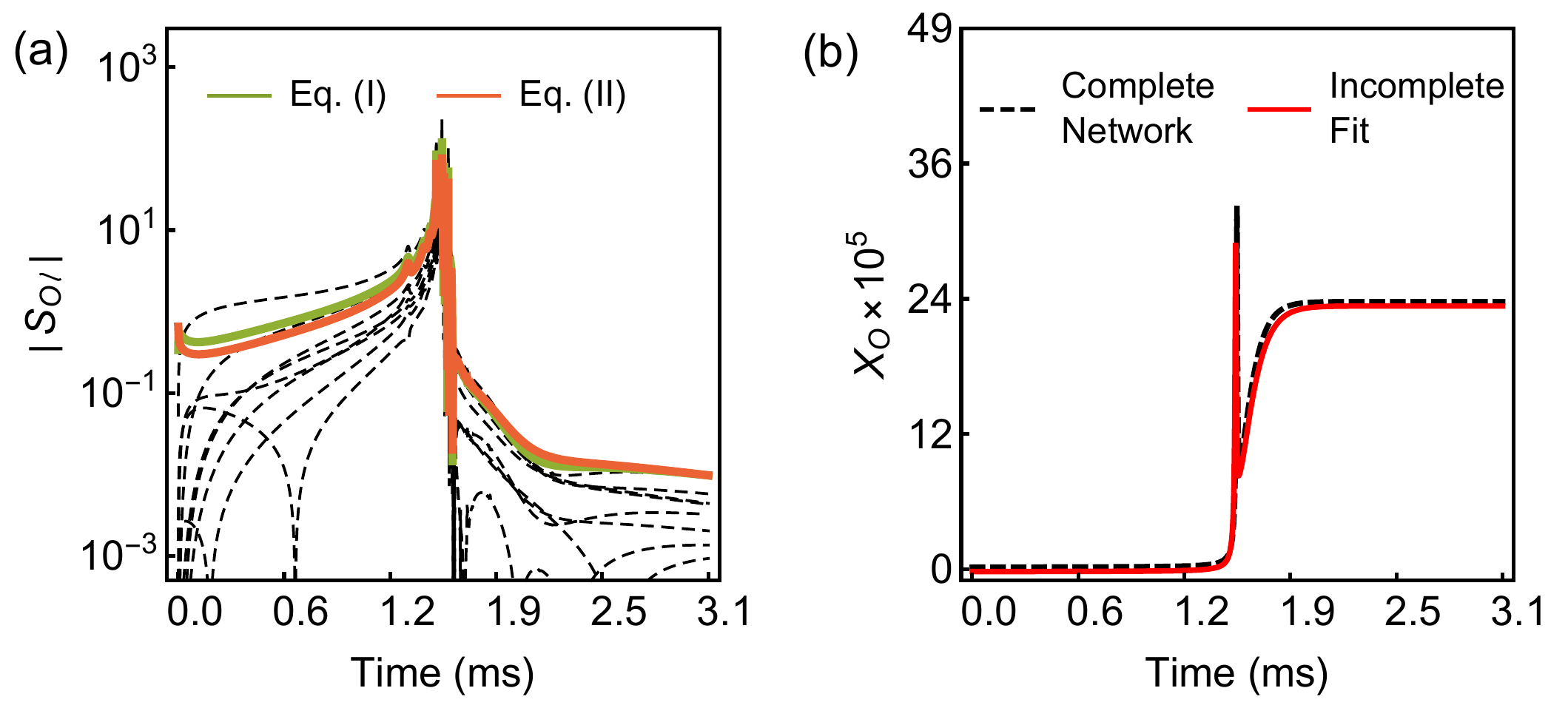}
\caption{{Simulated rate constant measurement for Eqs.~\eqref{eqn1} and \eqref{eqn2}. (a)  Local sensitivity $S_{\ce{O} \ell}$ vs time during ignition for Eq.~\eqref{eqn1} (green line), Eq.~\eqref{eqn2} (red line), and the other eight most sensitive reactions (dashed lines).  (b) Oxygen radical mole fraction $X_{\ce{O}}$ vs time during the ignition event corresponding to Experiment $14$ (see Table \ref{conditions} in Appendix \ref{acombustion}) for the complete network and a fitted incomplete network with $N_{\mathrm{ret}}=N_{\mathrm{rem}}=40$. The value of the rate constant for the fitted network provides an estimate for the rate constants $k_{\mathrm{est}}$.  \label{fig3} }}
\end{figure}

We perform a numerical optimization to minimize an error $\epsilon$ in order to fit the incomplete network to the measurements. The error we use quantifies the discrepancy in the peak oxygen radical concentration and ignition time, as described in Appendix~\ref{acombustion}. 
This particular quantity is employed in order to emulate the data available in real experiments, such as shock tube experiments based on oxygen radical concentration peaks \cite{2005_Baulch}.  
Furthermore, the reactions in Eqs.~\eqref{eqn1} and \eqref{eqn2} are known to be important for combustion modeling, and their local sensitivities $S_{\ce{O} \mathrm{I}}$ and $S_{\ce{O} \mathrm{II}}$ are among the largest of all the reactions in the network during the ignition events, as shown in Fig.~\ref{fig3}(a).  

\begin{figure}[t]
\includegraphics[width=\columnwidth]{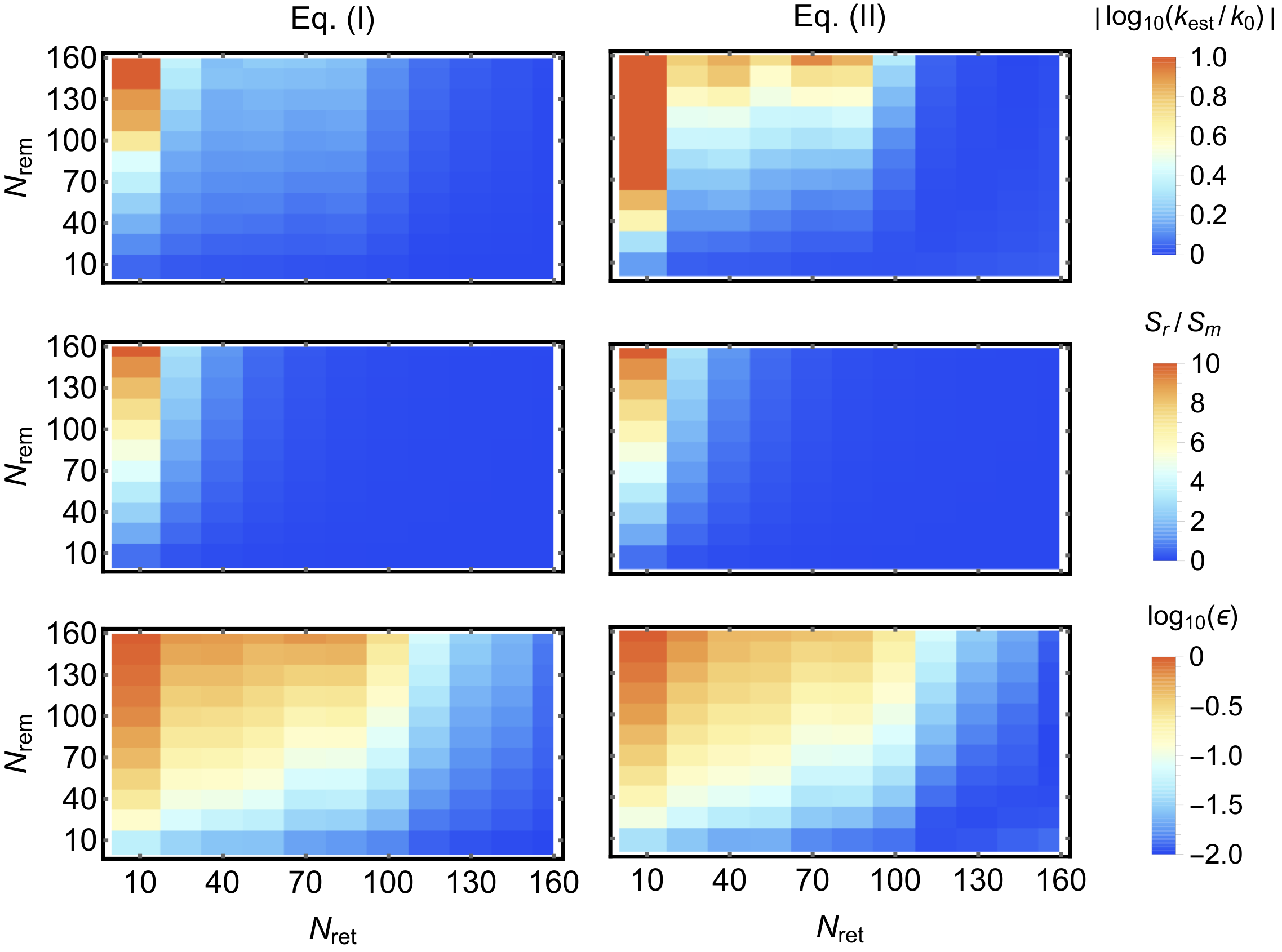}
\caption{{Average values of $\left \lvert \log_{10} \left({k_{\mathrm{est}}}/{k_0} \right) \right \rvert$ (top row), $S_{\mathrm{r}}/S_{\mathrm{m}}$ (center row), and $\log_{10} \left(\epsilon \right)$ (bottom row) as a function of $N_{\mathrm{rem}}$ and $N_{\mathrm{ret}}$ for Eq.~\eqref{eqn1} (left column) and Eq.~\eqref{eqn2} (right column). The averages follow similar trends, with magnitudes increasing for increasing $N_{\mathrm{rem}}$ and decreasing $N_{\mathrm{ret}}$. \label{fig4} }}
\end{figure} 

Incomplete networks are generated randomly by removing reactions with small oxygen sensitivity from the network. In order to quantify how the sensitivity of the removed reactions impacts the variability of the rate constant estimate, we retain from the $N_{\mathrm{tot}}$ total reactions a tunable number of reactions, $N_{\mathrm{ret}}$, with the largest $\ce{O}$-sensitivities, which are not candidates for removal in the generation of the incomplete networks.  Incomplete networks are then generated by randomly selecting a tunable number of reactions $N_{\mathrm{rem}}$ from the remaining $N_{\mathrm{tot}}-N_{\mathrm{ret}}$ reactions and removing them from the network.  For each value of $N_{\mathrm{ret}}$ and $N_{\mathrm{rem}}$, multiple incomplete networks are generated with differing random seeds, which allows us to statistically study the effects of missing information.  
Rate constant estimates are obtained from fits minimizing $\epsilon$. The fit of one sample realization is shown in Fig.~\ref{fig3}(b), but the quality of this fit varies significantly with each randomly realized incomplete network for fixed $N_{\mathrm{rem}}$ and $N_{\mathrm{ret}}$.  

Under any particular thermodynamic condition, minimizing the error results in a rate constant estimate $k_{\mathrm{est}}$ for Eq.~\eqref{eqn1} or Eq.~\eqref{eqn2}, but these estimates will vary with the thermodynamic conditions. For example, for one incomplete network generated with $N_{\mathrm{ret}}=40$ and $N_{\mathrm{rem}}=40$, the estimate for Eq.~\eqref{eqn1} varied considerably among the $27$ thermodynamic conditions described in Appendix \ref{acombustion}, with values ranging from $0.06$ to $4.71$ times as large as the actual rate constant $k_0$. 
Each estimate made in this fashion uses only a single observational data point, and so we may hope to eliminate the variation in the estimate by combining many observations in a single fit to produce an estimate. 

To test if measurements of multiple data points will improve the estimate, we performed $1000$ rate constant estimates for Eqs.~\eqref{eqn1} and \eqref{eqn2} for incomplete networks generated with $N_{\mathrm{rem}}$ and $ N_{\mathrm{ret}}$ varying from $10$ to $160$ by increments of $15$ (in total, $2.42\times10^5$ estimates), where each estimate was obtained by minimizing the aggregated error over all $27$ thermodynamic conditions. Figure \ref{fig4} shows how the average results from these simulations vary with $N_{\mathrm{rem}}$ and $N_{\mathrm{ret}}$. For each estimate, we consider the ratio of the sensitivities of the removed and measured reactions, $S_{\mathrm{r}}/S_{\mathrm{m}}$, the magnitude of logarithmic deviation of the rate constant estimate (representing the orders-of-magnitude difference between the rate constant estimate and the actual rate constant), $\left \lvert \log_{10} \left({k_{\mathrm{est}}}/{k_0} \right) \right \rvert$, and the logarithm of the error, $\log_{10} \left(\epsilon \right)$. 
It is clear from these results that after increasing the amount of observational data in the rate constant estimates, there can still be significant variation in the estimates even when relatively few of the least sensitive reactions are missing from the network and the fit appears adequate.
\begin{figure}[t]
\includegraphics[width=\columnwidth]{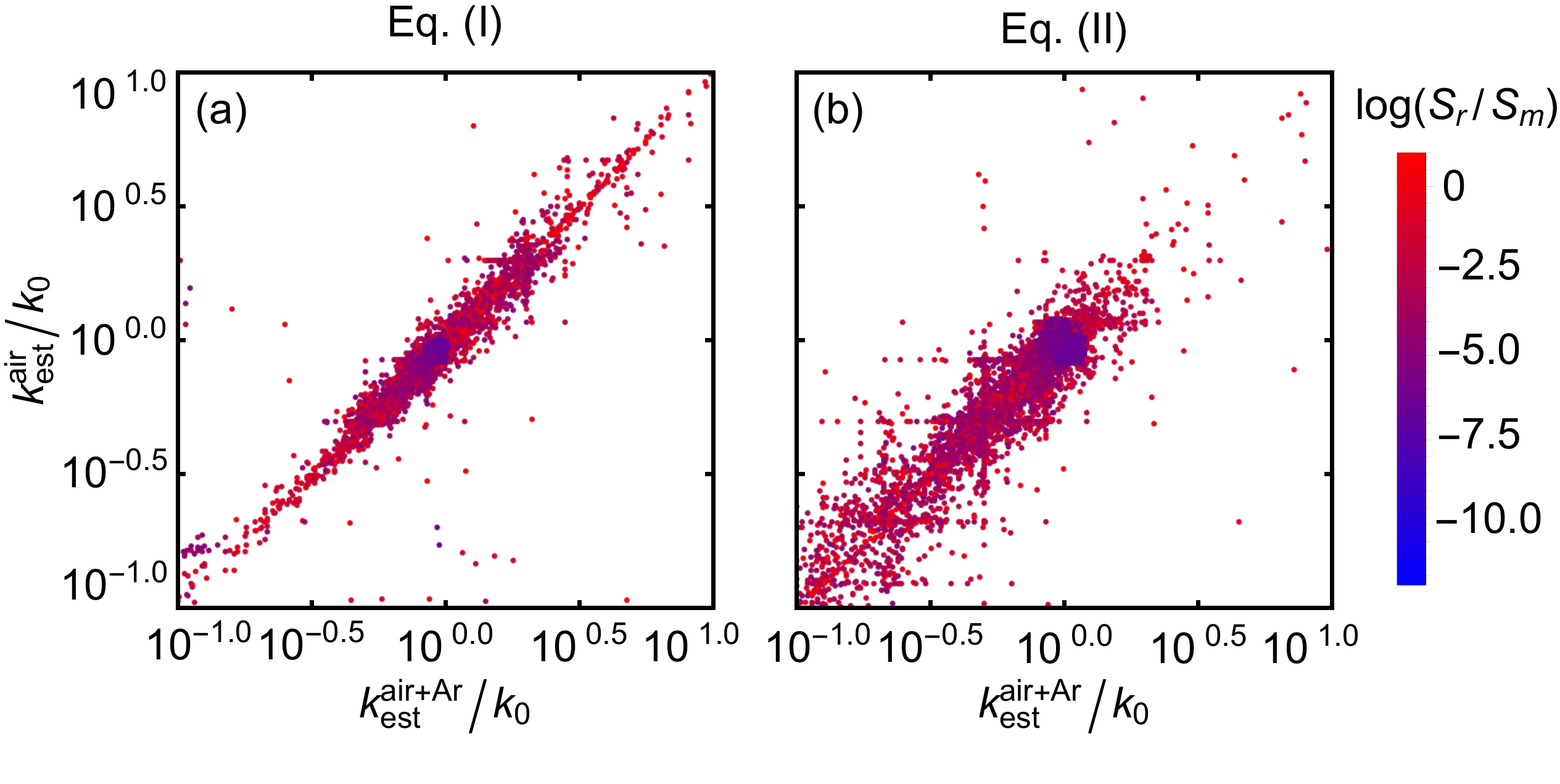}
\caption{Rate constant estimate with argon dilution vs rate constant estimate without argon dilution for (a) Eq.~\eqref{eqn1} and (b) Eq.~\eqref{eqn2}.  The same incomplete network is used in each pair of measurements, and the color code indicates the ratio of the sensitivities between the removed reactions and the measured reaction. \label{fig5}}
\end{figure}
\begin{figure*}
\includegraphics[width=2\columnwidth]{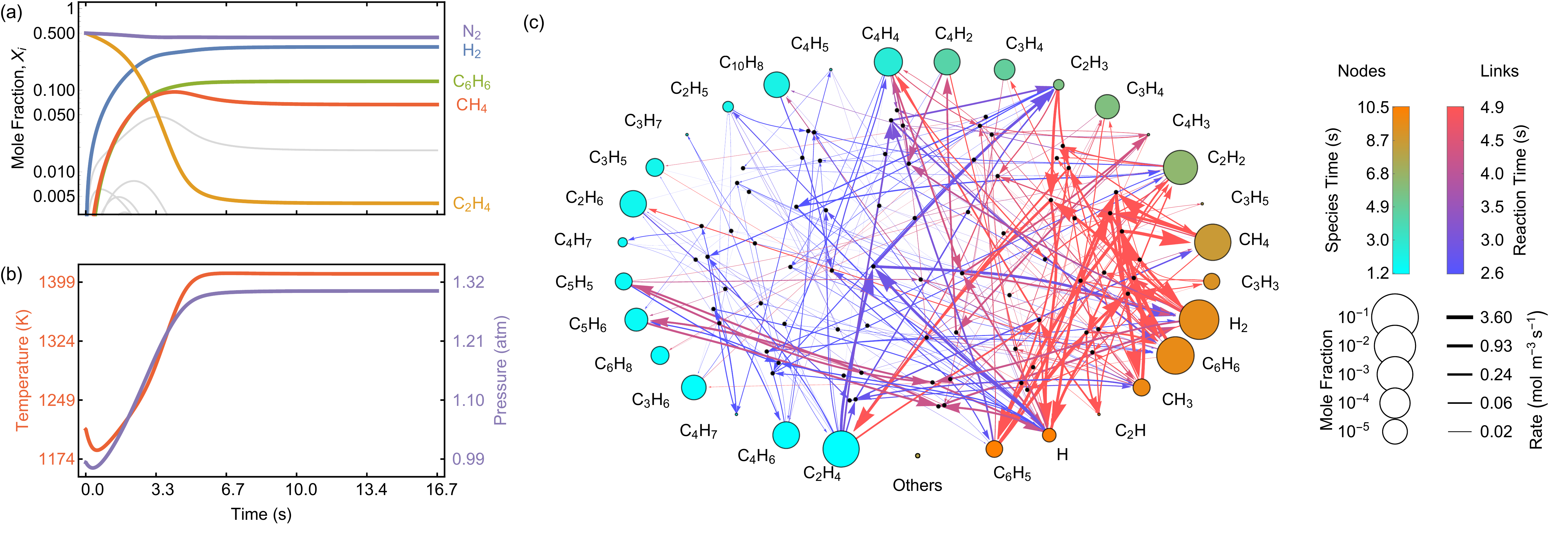}
\caption{Counterpart of Fig.~\ref{fig2} for ethylene pyrolysis. The ethylene fuel breaks down in the absence of oxygen in a slower and more temporally distributed process than in the case of combustion, where oxygen is present. 
\label{fig6}}
\end{figure*}

Having incorporated multiple thermodynamic conditions in each estimate, it is interesting to study how the estimates change under yet different conditions.  
We performed a second set of rate constant measurements using the same set of incomplete networks described above under slightly different experimental conditions by diluting the gas with about ten percent argon.  Argon, a neutral buffer which does not react chemically, is sometimes used in shock-tube experiments to adjust conditions, such as pressure, under the assumption that its presence will not alter the results.  However, besides deviating again from the true rate constants, the same incomplete networks also exhibit substantial differences between rate constant measurements performed in air with and without argon dilution, as shown in Fig.~\ref{fig5}. Among all the paired measurements, the Pearson correlation coefficient between rate constant estimates with no argon and with argon dilution was $0.93$ for Eq.~\eqref{eqn1} and $0.78$ for Eq.~\eqref{eqn2}, while the fraction of variance unexplained by a linear regression between the two conditions was, respectively,  $14\%$ and $39\%$.  Furthermore, while networks with small removal sensitivities tend to produce rate constant estimates closer to the real value, the scatter in Fig.~\ref{fig5} is not substantially smaller when the estimates are good or when the removal sensitivity is small. Thus, network incompleteness can clearly result in variable rate constant estimates under differing conditions even when the reactions being measured have large sensitivities compared to the neglected reactions and the change in conditions is seemingly innocuous.

\section{Ethylene pyrolysis}
\label{pyrolysis}
\begin{figure}[b]
\begin{center}
\includegraphics[width=\columnwidth]{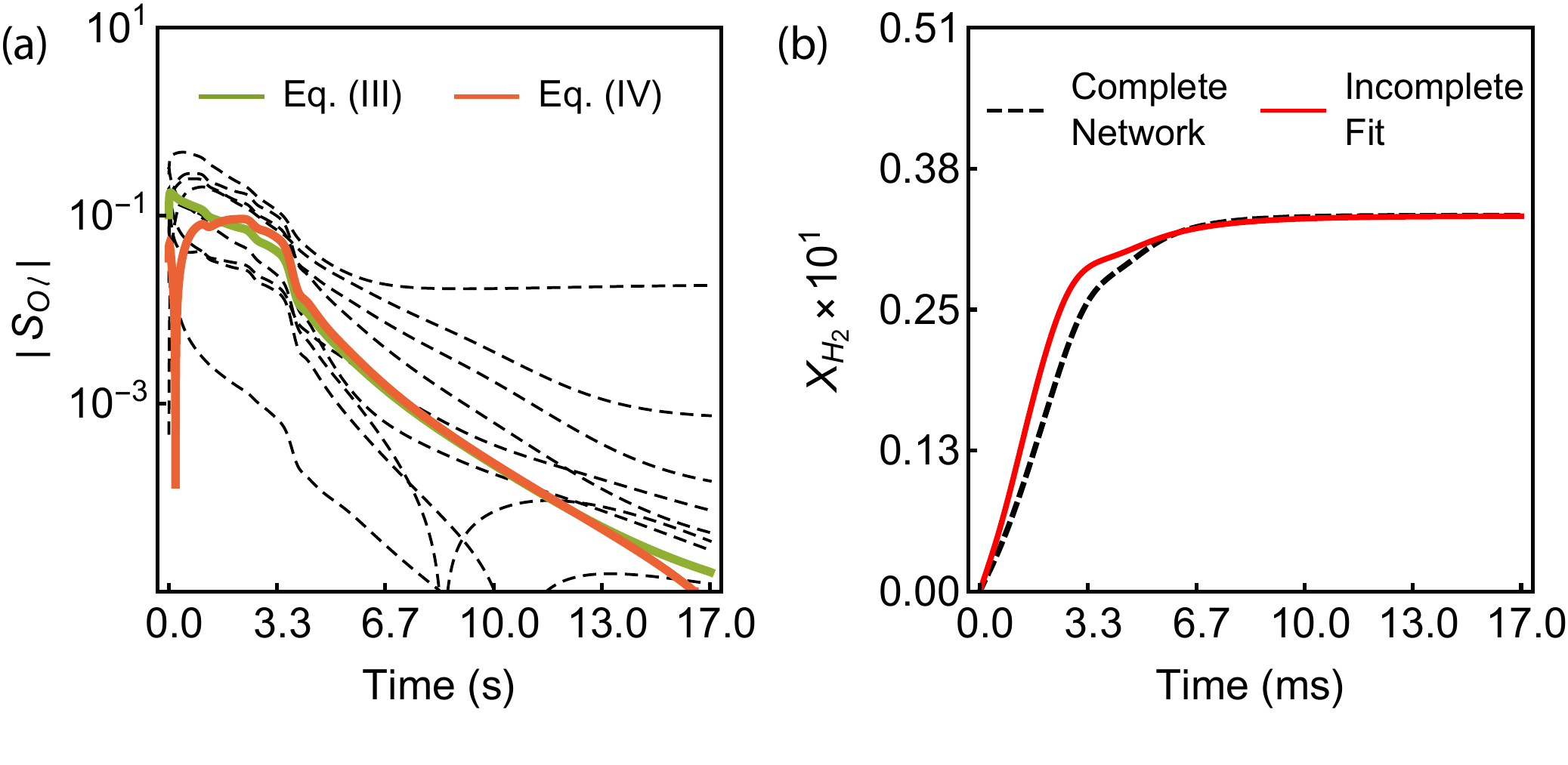}
\caption{{Counterpart of Fig.~\ref{fig3} for pyrolysis reactions in Eqs.~\eqref{eqn3}-\eqref{eqn4}, with $N_{\mathrm{rem}}=N_{\mathrm{ret}}=200$. Incomplete network models are fitted to the observed yields of $X^{\mathrm{o}}_{\ce{H2}}$ and $X^{\mathrm{o}}_{\ce{CH4}}$ to simulate measurements of the rate constants. \label{fig7}}}
\end{center}
\end{figure}
\begin{figure}[b]
\includegraphics[width=\columnwidth]{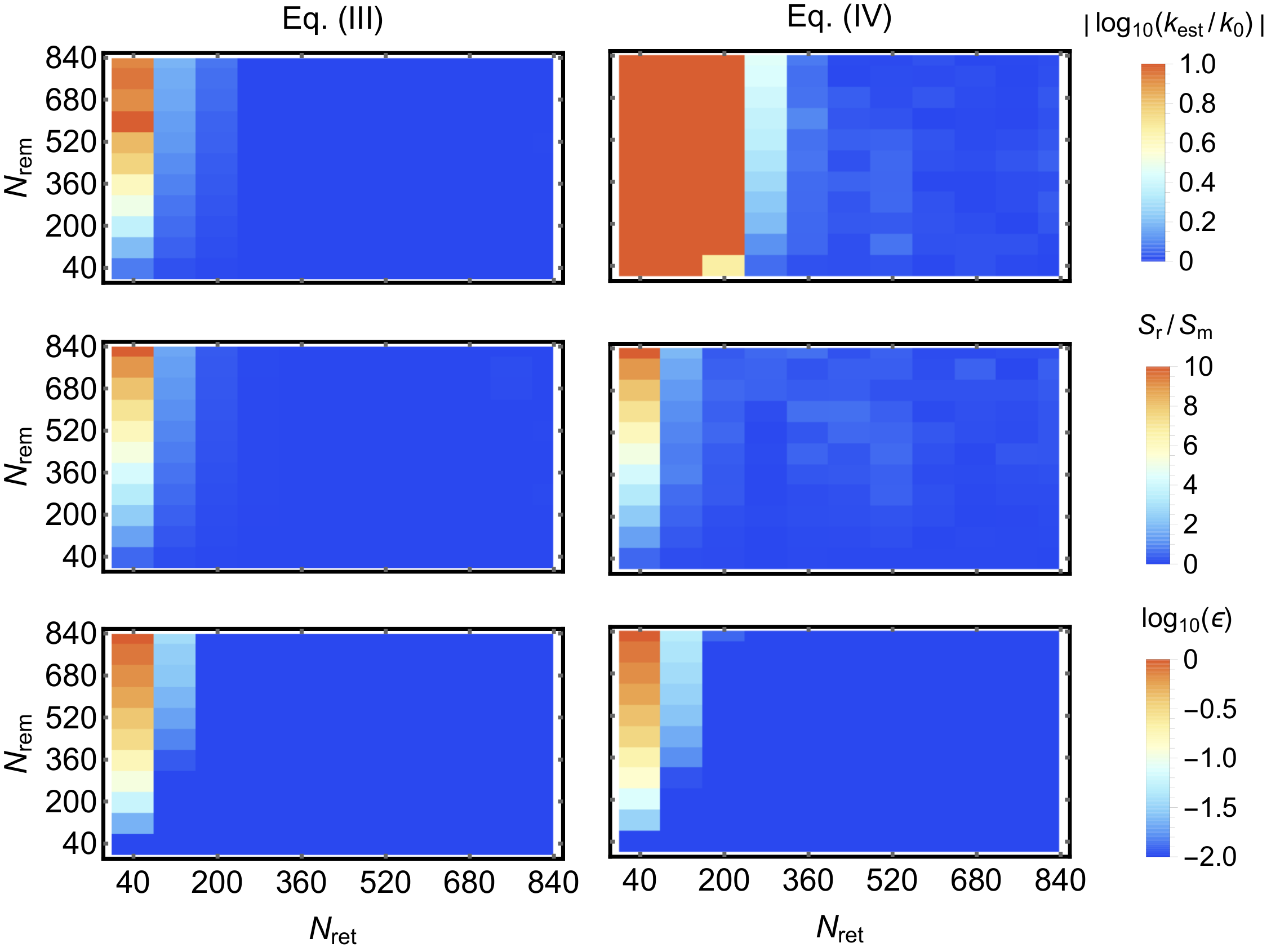}
\caption{{Counterpart of Fig.~\ref{fig4} for ethylene pyrolysis. The rate constant estimate $k_{\mathrm{est}}$, removal sensitivity $S_r$, and error $\epsilon$ follow the same general trends as for natural gas combustion, but estimates are generally better for Eq.~\eqref{eqn3} and worse for Eq.~\eqref{eqn4}. \label{fig8}}}
\end{figure}
To illustrate that these results generalize beyond combustion, we next consider a different complex reaction network describing ethylene pyrolysis. Pyrolysis is a common industrial chemical process involving the burning of organic compounds in the absence of oxygen.  In this process, larger organic molecules like $\ce{C2H4}$ (ethylene) break down into smaller molecules such as $\ce{H2}$ and $\ce{CH4}$.  We employ a recent model of light hydrocarbon pyrolysis consisting of $84$ species and $1698$ reactions as our complete pyrolysis network \cite{2012_Ranzi}. Figure \ref{fig6} shows the evolution of the process and network structure in the pyrolysis case, as in Fig.~\ref{fig2}.   

\begin{figure*}
\includegraphics[width=2\columnwidth]{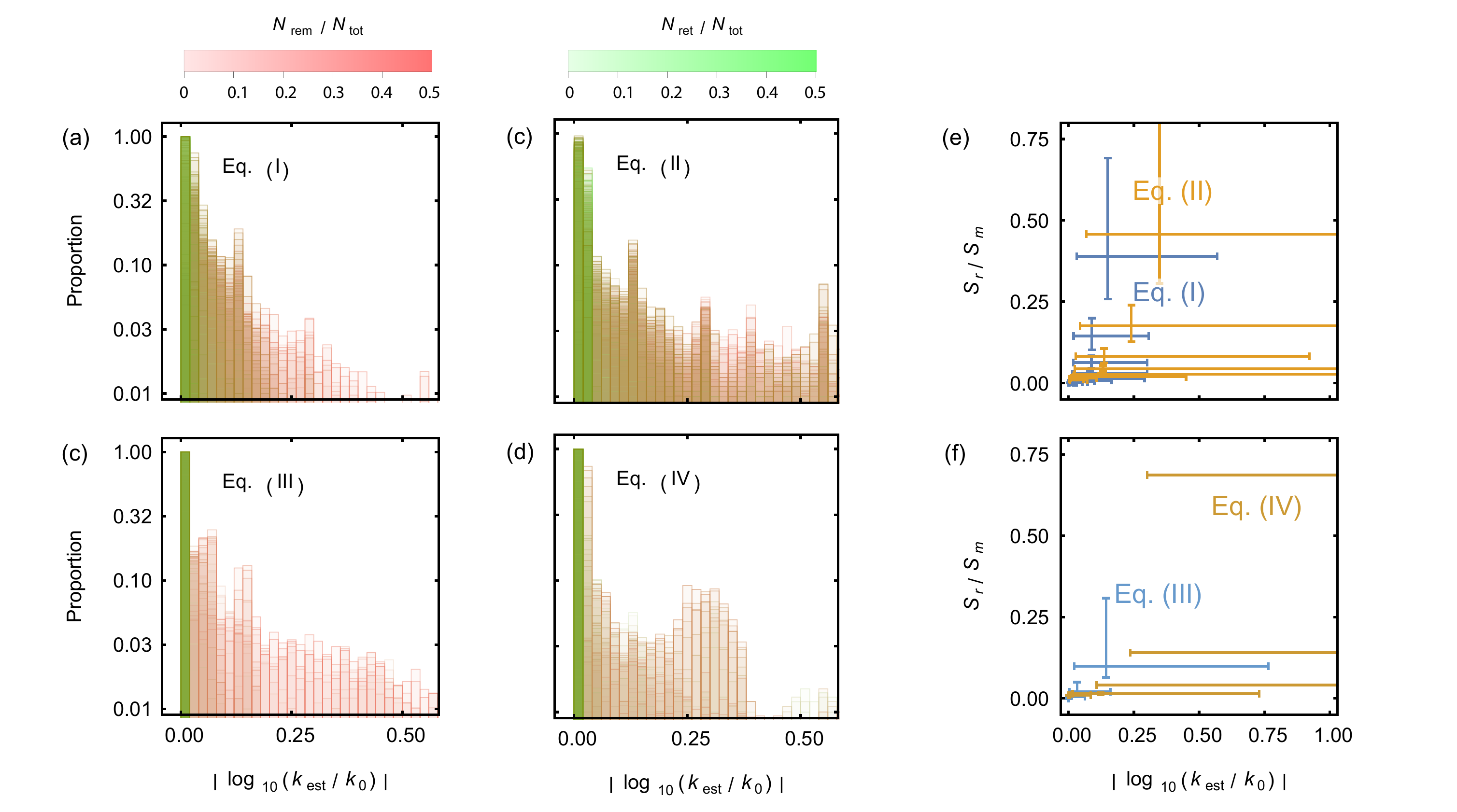}
\caption{Distributions of rate constant estimates for Eqs.~\eqref{eqn1}-\eqref{eqn4}. (a-d) Histograms of the rate constant estimates for Eqs.~\eqref{eqn1}-\eqref{eqn4}, with the red-green color blend indicating the proportion of removed and retained reactions for each set of histograms. (e-f) Relative sensitivity of the removed reactions vs rate constant estimates for Eqs.~\eqref{eqn1}-\eqref{eqn4}, with all data divided into $25$ evenly-occupied bins in each axis and with error bars showing the $90\%$ quantiles for each bin.  \label{fig9}}
\end{figure*}

For ethylene pyrolysis, we consider rate constant measurements based on net yields of $\ce{H2}$ and $\ce{CH4}$ rather than the peak $\ce{O}$ concentration and ignition time used previously.   
We allow the reaction to evolve until the $\ce{H2}$ concentration reaches its steady state with $27$ different thermodynamic conditions relevant to low pressure chemical vapor deposition.  We consider two reactions,
\begin{align}
\ce{H + C2H4 -> H2 + C2H3},  \label{eqn3} \tag{III} \\
\ce{C2H5 + C2H4 -> C2H6 + C2H3},  \label{eqn4} \tag{IV}
\end{align}
each of which has large sensitivity with respect to the $\ce{H2}$ and $\ce{CH4}$ net yields.  The  error $\epsilon$ was minimized as before to obtain the rate constant estimates.
Appendix~\ref{apyrolysis} describes details for these pyrolysis simulations. Figures \ref{fig7}-\ref{fig8} show counterparts to Fig.~\ref{fig3}-\ref{fig4} in the case of pyrolysis. It is clear from these results that, as was the case for methane combustion, rate constant estimates may vary considerably for ethylene pyrolysis when relatively few reactions are missing even if the missing reactions have small sensitivities and the fit errors are small.

\section{Combustion/Pyrolysis Comparisons}
\label{comparison}
\begin{table}[b]
\begin{tabular}{l  c  r  r  r  r}
$x$ & $y$ & ~\eqref{eqn1}~ & ~\eqref{eqn2}~ & ~\eqref{eqn3}~ & ~\eqref{eqn4}~ \\
\hline \phantom{.} & \phantom{.} & \phantom{.} & \phantom{.} &\phantom{.} &  \phantom{.} \\[0.05em]
${S_{\mathrm{r}}}/{S_{\mathrm{m}}}$ & $\log_{10}\left( \epsilon \right)$ & $0.48$ & $0.49$ & $0.59$ & $0.47$ \\
${S_{\mathrm{r}}}/{S_{\mathrm{m}}}$ & $\left \lvert \log_{10} \left({k_{\mathrm{est}}}/{k_0}\right) \right\rvert$  & $0.43$ & $0.46$ & $0.47$ & $0.45$ \\
$\left \lvert \log_{10} \left({k_{\mathrm{est}}}/{k_0}\right) \right\rvert$ & $\log_{10}\left( \epsilon \right)$  & $0.30$ & $0.38$ & $0.32$ & $0.64$ \\
\end{tabular}
\caption{Pearson correlation coefficients between quantities $x$ and $y$ for the distribution of simulated rate constant measurements in the combustion reactions in Eqs.~\eqref{eqn1} and \eqref{eqn2} and the pyrolysis reactions in Eqs.~\eqref{eqn3} and \eqref{eqn4}. \label{table2}}
\end{table}
\begin{figure}[b]
\includegraphics[width=\columnwidth]{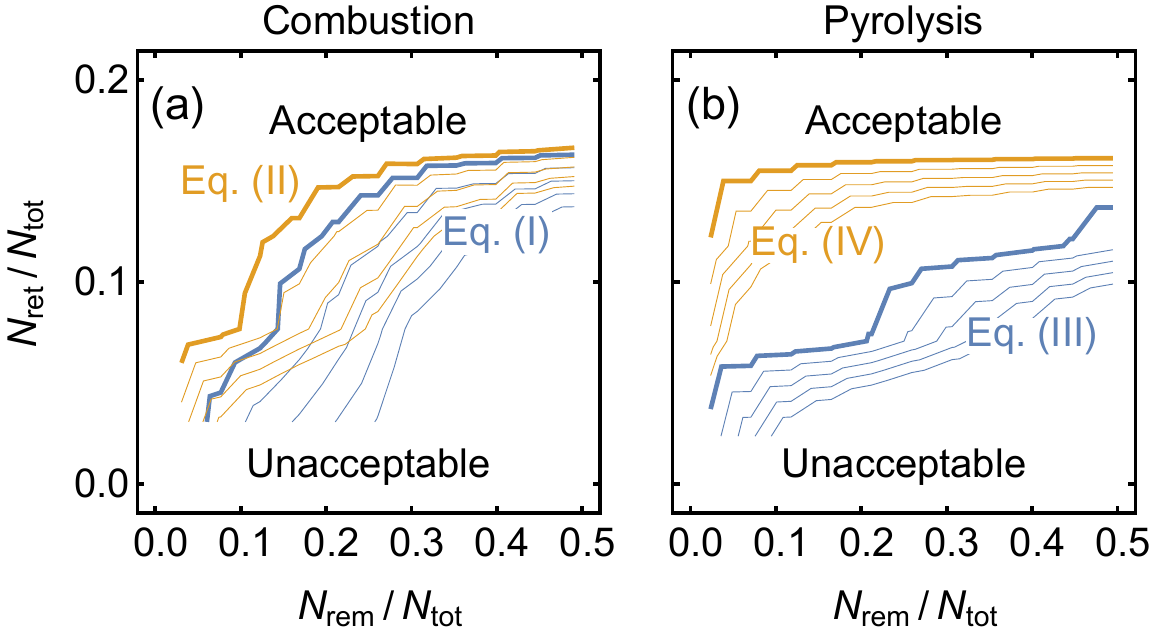}
\caption{{Acceptable and unacceptable regions of network completeness for (a) the combustion network and (b) the pyrolysis network. The rate constant estimates are deemed acceptable only if more than $95\%$ of incomplete network fits result in rate constant estimates which differ from the actual rate constant by less than a factor of two. 
Thin lines show how the boundary varies as the acceptable threshold is reduced to $90\%$,  $85\%$,  $75\%$, and $70\%$. }\label{fig10}}
\end{figure}
The distributions of rate constant estimates for the combustion and pyrolysis networks are broadly similar, as summarized by the histograms in Fig.~\ref{fig9}(a)-(d). The rate constant estimates typically deviate much more than the error or the removal sensitivities in all cases, despite the differing measurement targets and reaction networks. Table \ref{table2} shows the correlation coefficients between measurement quantities for each reaction in each process; there are moderate to weak correlations between all quantities.  Thus, in both these processes, the local sensitivity metric correlates weakly with the rate constant estimates and is not a very reliable measure of link importance. Even when the removed reactions have local sensitivity totaling to much less than the sensitivity of the measured links, the rate constant estimates can vary considerably, as shown in Fig.~\ref{fig9}(e)-(f).  In the networks we generated, for example, the smallest values of $S_r/S_m$ leading to order-of-magnitude changes with $|\log_{10} (k_{\mathrm{est}}/k_0)| > 1$ are just $0.005$ for Eq.~\eqref{eqn1}, $0.0046$ for Eq.~\eqref{eqn2}, $0.019$ for Eq.~\eqref{eqn3}, and $0.0024$ for Eq.~\eqref{eqn4}.  While computational costs prohibit us from testing all reactions in these networks, less systematic simulations suggest that other reactions typically exhibit even greater variations in rate constant estimates, as expected given that the reactions \eqref{eqn1}-\eqref{eqn4} are among the most sensitive ones in the networks.

To compare the two processes in more detail, we consider the effect of varying the number of removed and retained reactions in the simulated measurements. We divide the plane spanned by $N_{\mathrm{ret}}$ and $N_{\mathrm{rem}}$ into an acceptable region (in which more than $95\%$ of the rate constant estimates from fits of incomplete networks deviate from the actual value by less than a factor of two) and an unacceptable region (in which this is not the case).  Figure \ref{fig10} compares the acceptable regions in the combustion and pyrolysis cases by normalizing the axes by $N_{\mathrm{tot}}$, the total number of reactions in each network. 
In order to achieve acceptable measurements  with just $10\%$ of reactions missing for all cases, the retained reactions must amount to more than $15\%$ of the network.  While somewhat smaller $N_{\mathrm{ret}}/N_{\mathrm{tot}}$ is adequate for acceptable measurements for Eqs.~\eqref{eqn1}-\eqref{eqn3} in this case, the proportion of retained reactions required for acceptable measurements rises to about $15\%$ in all cases when $50\%$ of the reactions are missing.

Interestingly, the natural gas network shows a stronger dependence on the proportion of removed reactions than the pyrolysis network, indicating that the pyrolysis network has a greater proportion of reactions that are unimportant to the dynamics of the measurement. This may result from the more complex chain reaction involved in ignition compared to the slower and more temporally distributed pyrolysis process.

\section{Discussion}
\label{discussion}
Our results show that observed variations in rate constants of combustion and pyrolysis networks can be largely attributed to the use of incomplete network models that are missing reactions that were presumed to be negligible. We demonstrated that even when the neglected reactions have small cumulative sensitivity compared to the sensitivity of the measured reaction, rate constant estimates in incomplete networks can vary by one order of magnitude. Furthermore, rate constant estimate deviations and local sensitivity analysis showed weak statistical correlations. While these results highlight significant challenges in constructing a reliable and transferable kinetic model for these important chemical processes, they also reveal that current observations of variability in rate constant estimates are not necessarily an indication that a complete network model applicable across all conditions is unattainable.  Yet, since measurements of individual rate constants using the current incomplete network models cannot produce reliable results, iteratively refining individual parameters through targeted experiments is unlikely to ever produce accurate estimates.  It would therefore be desirable to accumulate large data sets from measurements across all relevant conditions in order to simultaneously estimate all rate constants in the model. 

Our results also have implications for other processes described by complex chemical reaction networks, including atmospheric processes \cite{2008_Atkinson,runge2015identifying, 2016_Debiagi} as well as biochemical processes, such as those mediated by metabolic networks \cite{2013_Chakrabarti, 2014_Khodayari,2014_Link}. Compared to the combustion and pyrolysis networks considered here, parameter estimation in metabolic networks in particular is still in its infancy and could benefit significantly from more informed formulations at the outset. One promising direction for improvement is suggested by information-geometric and sloppy modeling approaches \cite{2014_Transtrum, 2015_Transtrum, 2016_White}.  Using such methods, it may be possible to identify measurement targets as algebraic combinations of rate constants that are more robust to the variations caused by missing network links. Another promising information theoretic approach for predicting network parameters from data would be to employ maximum entropy models, whose degeneracy issues have been recently addressed \cite{2015_Horvat}. Alternatively, it may be possible to directly estimate reaction fluxes rather than species concentrations in order to disentangle reactions.   In principle, molecular dynamics simulations can be used to estimate fluxes even if they are experimentally inaccessible, provided that the molecular force fields can be modeled realistically \cite{2015_Dontgen,2017_Alaghemandi}.

Beyond chemical reaction networks, social networks \cite{2004_Eubank,2007_Liben, 2018_Jorgensen} and  infrastructure networks may also benefit from our findings.   In power grids \cite{2013_Backhaus, 2013_motter}, for example, the network components are assumed to be  known but  can involve uncertainties and the network itself may also vary, as not all components are necessarily operational at all times. Our results suggest that power-grid models may become unreliable even if the component parameters are accurate unless they are updated in real time to account for missing components.   In other applications, our results also emphasize an important trade-off between model accuracy and parameter accuracy when dimension reduction is implemented by network trimming. That is, trimmed networks may lead to accurate results under specific conditions at the cost of requiring parameter values that differ significantly from their actual values and that may change substantially as conditions change. We anticipate that the impact of missing information on parameter estimation will ultimately depend on the class of networks under consideration, much in the same way as optimal link detection varies across scientific domains \cite{2020_Ghasemian}.

\acknowledgments{The authors thank Igor Mezi\'{c} for insightful discussions. This work was supported by MURI Grant No.\ W911NF-14-1-0359 and ARO Grant No.\ W911NF-19-1-0383.}

\begin{appendix}
\section{Electric circuit analysis}
\label{acircuit}
We first derive the measured currents for the circuit Fig.~\ref{fig1} using Kirchhoff's laws. The voltage across the blue resistor is known, so the current moving from left to right across the blue resistor is $I_3 = (V_{\mathrm{adj}}-V_1)/R_1$. Since the black nodes are neither sources nor sinks of current, the current from top to bottom across the red resistor must be $I_2+I_3 = ( V_{\mathrm{adj}} - V_3)/R_2$. This same current must pass through the horizontal $1\Omega$ resistor, so that $I_2+I_3 = V_3/(1\Omega)$.  It follows that $V_3 = V_{\mathrm{adj}}/(R_2+1\Omega)$ and $I_2 = V_{\mathrm{adj}}/(R_2+1\Omega) - (V_{\mathrm{adj}}-V_{1})/R_1$. Similarly, the current across the vertical $1 \Omega$ resistor must be $I_1-I_3 = V_1 / (1\Omega)$, and thus $I_1 = V_1/(1\Omega) + (V_{\mathrm{adj}}-V_1)/R_1$.

Next, we consider the model of the experimenter who neglects the red resistor. According to this model, all the current drawn through the adjustable voltage source travels across the blue resistor, so that the estimated current is $I_2' = -(V_{\mathrm{adj}}-V_{1})/R_{\mathrm{est}}$, where $R_{\mathrm{est}}$ is the estimate of the resistance of the blue resistor. Similarly, the experimenter estimates the current drawn from the left voltage source to be  $I_1' = V_1/(1\Omega) + (V_{\mathrm{adj}}-V_1)/R_{\mathrm{est}}$. The square error for the incomplete model, $\epsilon^2 \equiv (I_1'-I_1)^2 + (I_2'-I_2)^2$, is thus
\begin{align}
\epsilon^2 &= \Delta^2  + \left(\frac{V_{\mathrm{adj}}}{R_2+1\Omega} + \Delta \right)^2,
\end{align}
where $\Delta=({V_{\mathrm{adj}}-V_1})(1/{R_{\mathrm{est}}}-1/{R_1})$. The least square estimate for the resistance in Fig.~\ref{fig1}(b) follows by minimizing $\epsilon^2$ with respect to $R_{\mathrm{est}}$. The local sensitivity $S_{ij}\equiv({R_j}/{I_i})({\partial I_i}/{\partial R_j})$ quantifies how the $i$th current varies with the $j$th resistor.  It seems intuitive that the least square estimate derived from the model that omits $R_2$ can become unreliable when $S_{i2}$ becomes large compared to $S_{i1}$. Indeed, for this circuit, the ratio $S_{22}/S_{21}=(R_2+1\Omega)^2(V_{\mathrm{adj}}-V_1)/R_1R_2V_{\mathrm{adj}}$ approaches zero as $V_{\mathrm{adj}}$ approaches $V_1$, where the estimate in Fig.~\ref{fig1}(b) becomes unreliable.
\renewcommand{\thetable}{C\arabic{table}}
\setcounter{table}{0}
\begin{table*}[t]
\begin{tabular}{| c | c | c | c | c | c | c |}
\hline & & & & & & \\
 \hspace{1em} Exp. \# \hspace{1em} & \hspace{1em} $T (\si{K}) $ \hspace{1em} & \hspace{1em} $P (\si{atm})$ \hspace{1em} & \hspace{1em} $\phi$  \hspace{1em} & \hspace{2em} $t^\mathrm{o}_{\mathrm{ig}} (\si{s})$ \hspace{2em} & \hspace{2em} $X^\mathrm{o}_{\ce{O}_\mathrm{peak}}$ \hspace{2em} &  \hspace{1em} $k_{\mathrm{est}}/k_0$ \hspace{1em} \\[0.5em]
\hline & & & & & & \\
$1$ & $1300$ & $0.1$ & $0.5$ &  $6.3\times10^{-2}$ & $1.5\times10^{-2}$ & $1.02$ \\
$2$ & $1300$ & $0.1$ & $1.0$ &  $9.0\times10^{-2}$ & $2.7\times10^{-4}$ & $1.01$ \\
$3$ & $1300$ & $0.1$ & $2.0$ &  $1.4\times10^{-1}$ & $1.8\times10^{-5}$ & $4.71$ \\
$4$ & $1300$ & $1.0$ & $0.5$ &  $1.2\times10^{-2}$ & $1.3\times10^{-2}$ & $1.03$\\
$5$ & $1300$ & $1.0$ & $1.0$ &  $1.7\times10^{-2}$ & $7.8\times10^{-5}$ & $1.09$ \\
$6$ & $1300$ & $1.0$ & $2.0$ &  $2.5\times10^{-2}$ & $1.0\times10^{-5}$ & $2.27$ \\
$7$ & $1300$ & $5.0$ & $0.5$ &  $2.9\times10^{-3}$ & $7.4\times10^{-3}$ & $1.11$ \\
$8$ & $1300$ & $5.0$ & $1.0$ &  $4.1\times10^{-3}$ & $7.6\times10^{-5}$ & $1.30$ \\
$9$ & $1300$ & $5.0$ & $2.0$ &  $6.3\times10^{-3}$ & $6.8\times10^{-6}$ & $0.30$ \\

$10$ & $1500$ & $0.1$ & $0.5$ &  $7.6\times10^{-3}$ & $1.7\times10^{-2}$ & $1.03$ \\
$11$ & $1500$ & $0.1$ & $1.0$ &  $9.0\times10^{-3}$ & $4.2\times10^{-4}$ & $1.02$ \\
$12$ & $1500$ & $0.1$ & $2.0$ &  $1.2\times10^{-2}$ & $3.3\times10^{-5}$ & $2.00$ \\
$13$ & $1500$ & $1.0$ & $0.5$ &  $1.1\times10^{-3}$ & $1.5\times10^{-2}$ & $1.04$ \\
$14$ & $1500$ & $1.0$ & $1.0$ &  $1.5\times10^{-3}$ & $3.2\times10^{-4}$ & $1.03$ \\
$15$ & $1500$ & $1.0$ & $2.0$ &  $2.2\times10^{-3}$ & $2.1\times10^{-5}$ & $0.06$ \\
$16$ & $1500$ & $5.0$ & $0.5$ &  $3.3\times10^{-4}$ & $1.1\times10^{-2}$ & $1.07$ \\
$17$ & $1500$ & $5.0$ & $1.0$ &  $4.5\times10^{-4}$ & $2.0\times10^{-4}$ & $1.09$ \\
$18$ & $1500$ & $5.0$ & $2.0$ &  $6.6\times10^{-4}$ & $1.5\times10^{-5}$ & $0.14$ \\

$19$ & $1700$ & $0.1$ & $0.5$ &  $1.9\times10^{-3}$ & $1.8\times10^{-2}$ & $1.03$ \\
$20$ & $1700$ & $0.1$ & $1.0$ &  $1.9\times10^{-3}$ & $5.9\times10^{-4}$ & $1.05$ \\
$21$ & $1700$ & $0.1$ & $2.0$ &  $2.2\times10^{-3}$ & $5.4\times10^{-5}$ & $2.17$ \\
$22$ & $1700$ & $1.0$ & $0.5$ &  $2.2\times10^{-4}$ & $1.6\times10^{-2}$ & $1.05$ \\
$23$ & $1700$ & $1.0$ & $1.0$ &  $2.5\times10^{-4}$ & $5.0\times10^{-4}$ & $1.05$ \\
$24$ & $1700$ & $1.0$ & $2.0$ &  $3.2\times10^{-4}$ & $4.0\times10^{-5}$ & $1.87$ \\
$25$ & $1700$ & $5.0$ & $0.5$ &  $5.7\times10^{-5}$ & $1.3\times10^{-2}$ & $1.08$ \\
$26$ & $1700$ & $5.0$ & $1.0$ &  $6.9\times10^{-5}$ & $3.5\times10^{-4}$ & $1.10$ \\
$27$ & $1700$ & $5.0$ & $2.0$ &  $9.4\times10^{-5}$ & $2.9\times10^{-5}$ & $0.11$ \\
\hline
\end{tabular}
\caption{Simulated experimental observations for natural gas combustion.  The temperature $T$, pressure $P$, and the fuel to air ratio $\phi$ are varied, resulting in different observed ignition times $t^\mathrm{o}_{\mathrm{ig}}$ and peak $\ce{O}$ radical mole fraction $X^\mathrm{o}_{\ce{O}_\mathrm{peak}}$. An example of the estimated rate constant $k_{\mathrm{est}}/k_0$  for fits of the corresponding observation for a network with $40$ reactions removed is shown in the last column. \label{conditions} }
\end{table*}

\section{Kinetic models and reaction sensitivity}
\label{akinetic}
The chemical reaction networks we consider are described by state variables corresponding to the molar fraction $X_{\ce{A_i}}$ of the $i$th species with name $\ce{A_i}$ in an ideal gas and thermodynamic variables including the temperature $T$ and pressure $P$, with volume held fixed and under adiabatic conditions. For simplicity, we consider well-mixed systems, so that there is no spatial dependence on variables.  Molar fractions vary because of elementary chemical reactions between species, such as the $\ell$th reaction  $\ce{ \sum_i \nu_{i \ell} A_{i} -> \sum_j \eta_{j_\ell} A_j}$, which is assumed to occur at a rate governed by mass-action kinetics (i.e., proportional to a rate constant $\kappa_\ell$ times the product of the reactants molar concentrations raised to their stoichiometric coefficients).  Each molar fraction increases at a rate given by the reactions that produce it and decreases at a rate given by the reactions that consume it, as ${{\mathrm d}X_{A_i}}/{{\mathrm d}t} = \sum_\ell \kappa_\ell \left( \eta_{i \ell} - \nu_{i \ell} \right)\prod_{j} X_{A_j}^{\nu_{j \ell}}$.  The energy released or consumed from each reaction (determined from thermodynamic data about the species) is converted to heat, which enters an energy conservation equation governing the evolution of the temperature, and the pressure and temperature are related through a constitutive relation, which we take as the ideal gas law.  Each rate constant has temperature dependence which is based on a modified Arrhenius equation $\kappa = k T^b \exp(-E/RT)$ where $k$, $b$, and $E$ are model parameters. Some three-body reactions also depend on pressure with a Troe falloff, and reversible reactions have rate constants derived from detailed balance.  The local sensitivity, $S_{i\ell} \equiv \kappa_\ell/X_i\times\partial X_i/\partial \kappa_\ell$, quantifies how  the state variable $X_i$ changes as the parameter $\kappa_\ell$ varies \cite{1990_Turanyi}. Small changes in the rate constants for reactions with large sensitivities produce large changes in the species molar fractions, indicating that sensitivity may be a useful metric for reaction importance.

\renewcommand{\thetable}{D\arabic{table}}
\setcounter{table}{0}
\begin{table*}[t]
\begin{center}
\begin{tabular}{| c | c | c | c | c | c | c| }
\hline & & & & & & \\[0.1em]
\hspace{1em} Exp. \# \hspace{1em} & \hspace{1em} $T (\si{K}) $ \hspace{1em} & \hspace{1em} $P (\si{atm})$ \hspace{1em} & \hspace{1em} $\phi$  \hspace{1em} & \hspace{2em} $X^\mathrm{o}_{\ce{H2}}$ \hspace{2em} & \hspace{2em} $X^\mathrm{o}_{\ce{CH4}}$ \hspace{2em} &  \hspace{1em} $k_{\mathrm{est}}/k_0$ \hspace{1em} \\[0.5em]
\hline & & & & & & \\[0.1em]
$1$ & $1000$ & $0.01$ & $0.5$ &  $2.8\times10^{-1}$ & $1.2\times10^{-2}$ & 0.93 \\
$2$ & $1000$ & $0.10$ & $0.5$ &  $2.6\times10^{-1}$ & $3.2\times10^{-2}$ & 0.96 \\
$3$ & $1000$ & $1.00$ & $0.5$ &  $2.1\times10^{-1}$ & $6.5\times10^{-2}$ & 0.95 \\
$4$ & $1000$ & $0.01$ & $1.0$ &  $4.0\times10^{-1}$ & $1.9\times10^{-2}$ & 0.93 \\
$5$ & $1000$ & $0.10$ & $1.0$ &  $3.7\times10^{-1}$ & $4.7\times10^{-2}$ & 0.95\\
$6$ & $1000$ & $1.00$ & $1.0$ &  $3.1\times10^{-1}$ & $9.5\times10^{-2}$ & 0.93 \\
$7$ & $1000$ & $0.01$ & $2.0$ &  $5.1\times10^{-1}$ & $2.6\times10^{-2}$ & 0.93 \\
$8$ & $1000$ & $0.10$ & $2.0$ &  $4.7\times10^{-1}$ & $6.2\times10^{-2}$ & 0.94 \\
$9$ & $1000$ & $1.00$ & $2.0$ &  $4.0\times10^{-1}$ & $1.2\times10^{-1}$ & 0.91 \\

$10$ & $1100$ & $0.01$ & $0.5$ &  $2.8\times10^{-1}$ & $1.0\times10^{-2}$ & 0.95 \\
$11$ & $1100$ & $0.10$ & $0.5$ &  $2.7\times10^{-1}$ & $2.4\times10^{-2}$ & 0.94 \\
$12$ & $1100$ & $1.00$ & $0.5$ &  $2.3\times10^{-1}$ & $5.4\times10^{-2}$ & 0.93 \\
$13$ & $1100$ & $0.01$ & $1.0$ &  $4.0\times10^{-1}$ & $1.7\times10^{-2}$ & 1.04 \\
$14$ & $1100$ & $0.10$ & $1.0$ &  $3.8\times10^{-1}$ & $3.7\times10^{-2}$ & 0.93 \\
$15$ & $1100$ & $1.00$ & $1.0$ &  $3.3\times10^{-1}$ & $8.0\times10^{-2}$ & 0.92 \\
$16$ & $1100$ & $0.01$ & $2.0$ &  $5.1\times10^{-1}$ & $2.3\times10^{-2}$ & 1.05 \\
$17$ & $1100$ & $0.10$ & $2.0$ &  $4.8\times10^{-1}$ & $5.1\times10^{-2}$ & 0.93 \\
$18$ & $1100$ & $1.00$ & $2.0$ &  $4.2\times10^{-1}$ & $1.1\times10^{-1}$ & 0.91 \\

$19$ & $1200$ & $0.01$ & $0.5$ &  $2.8\times10^{-1}$ & $8.9\times10^{-3}$ & 0.92 \\
$20$ & $1200$ & $0.10$ & $0.5$ &  $2.7\times10^{-1}$ & $2.0\times10^{-2}$ & 0.98 \\
$21$ & $1200$ & $1.00$ & $0.5$ &  $2.4\times10^{-1}$ & $4.4\times10^{-2}$ & 0.93 \\
$22$ & $1200$ & $0.01$ & $1.0$ &  $4.0\times10^{-1}$ & $1.5\times10^{-2}$ & 1.30 \\
$23$ & $1200$ & $0.10$ & $1.0$ &  $3.8\times10^{-1}$ & $3.1\times10^{-2}$ & 0.97 \\
$24$ & $1200$ & $1.00$ & $1.0$ &  $3.4\times10^{-1}$ & $6.7\times10^{-2}$ & 0.92 \\
$25$ & $1200$ & $0.01$ & $2.0$ &  $5.0\times10^{-1}$ & $2.1\times10^{-2}$ & 1.43 \\
$26$ & $1200$ & $0.10$ & $2.0$ &  $4.8\times10^{-1}$ & $4.3\times10^{-2}$ & 0.97 \\
$27$ & $1200$ & $1.00$ & $2.0$ &  $4.3\times10^{-1}$ & $8.9\times10^{-2}$ & 0.91 \\
\hline
\end{tabular}
\end{center}
\caption{Simulated experimental observations for ethylene pyrolysis. The temperature $T$, pressure $P$, and fuel purity $\phi$ are varied, resulting in different net yields $X^\mathrm{o}_{\ce{H2}}$ and $X^\mathrm{o}_{\ce{CH4}}$. An example of the estimated rate constant $k_{\mathrm{est}}/k_0$  for fits of the corresponding observation for a network with $200$ reactions removed is shown in the last column. \label{conditions2}}
\end{table*}

\section{Simulated combustion measurements}
\label{acombustion}
We consider $27$ thermodynamic conditions for our measurement simulations that evenly span the range of thermodynamic values over which the GRI network was designed to model ignition, as listed in Table \ref{conditions}.  The initial concentrations in the simulation are a mixture of air and methane, with 8 parts $\ce{N2}$, 2 parts $\ce{O2}$, and $\phi$ parts fuel $\ce{CH4}$.  For the measurements in air diluted with argon, an additional $1$ part $\ce{Ar}$ was added to the initial conditions.

Experiments cannot generally observe the concentration of all species or reaction fluxes during the course of ignition.  Instead, specific targets are identified to measure and fit.  For this study, we consider the ignition time $t_{\mathrm{ig}}$ and the peak in the oxygen radical mole fraction $X_{\ce{O}_\mathrm{peak}}$ that occurs at ignition, where $t_{\mathrm{ig}}$ is taken as the time of the oxygen peak.   For simplicity, we fix the values of $b$ and $E$ in the rate constants to their values in the original network and optimize only over the pre-exponential factor $k$. We optimize over the error quantity
\begin{equation}
\epsilon^2 \equiv  {\left(\frac{t^\mathrm{o}_{\mathrm{ig}}-t^\mathrm{f}_{\mathrm{ig}}}{t^\mathrm{o}_{\mathrm{ig}}}\right)^2+\left(\frac{X^\mathrm{o}_{\ce{O}_\mathrm{peak}}-X^\mathrm{f}_{\ce{O}_\mathrm{peak}}}{X^\mathrm{o}_{\ce{O}_\mathrm{peak}}}\right)^2}, \label{error}
\end{equation}
where the superscripts $\mathrm{o}$ and $\mathrm{f}$ indicate the observed and fitted values, respectively.  
When the observation from multiple thermodynamic conditions are combined to produce a rate constant estimate, the square errors are totaled to give the aggregated error. 
The particular optimization target in Eq.~\eqref{error} is intended to emulate choices made in real combustion experiments \cite{2005_Baulch}. 

We use a Brent optimization algorithm, with a bracket found from an initial interval of $[k_0/2, 2k_0]$, where $k_0$ is the original network's rate constant value. To simulate dynamics, we used the open-source software Cantera to numerically integrate the coupled ordinary differential equations in the combustion and pyrolysis networks. The $2$-norm (over the time indexes) of the local sensitivity $S_{\ce{O} \ell}$ (and $S_{\ce{H2} \ell}$ in the case of pyrolysis below) was calculated for all reactions and averaged over all experimental conditions to rank reactions by their sensitivity.  The removal sensitivity $S_{\mathrm{r}}$ is the total sensitivity of all reactions removed to generate the incomplete network, and the measurement sensitivity $S_{\mathrm{m}}$ is the sensitivity of the reaction whose rate constant is being estimated.  

\section{Simulated pyrolysis measurements}
\label{apyrolysis}
Table \ref{conditions2} shows the experimental conditions for the pyrolysis rate constant measurements.  The fuel purity $\phi$ in this case indicates the ratio of ethylene to $\ce{N2}$ in the initial condition. These conditions were chosen for their applicability to low pressure chemical vapor deposition.  For ethylene pyrolysis, we optimize over the measurement error
\begin{equation}
\epsilon^2 \equiv {\left(\frac{X^\mathrm{o}_{{\ce{CH4}}}-X^\mathrm{f}_{{\ce{CH4}}}}{X^\mathrm{o}_{{\ce{CH4}}}}\right)^2+\left(\frac{X^\mathrm{o}_{{\ce{H2}}}-X^\mathrm{f}_{{\ce{H2}}}}{X^\mathrm{o}_{{\ce{H2}}}}\right)^2}, \label{error2}
\end{equation}
where the yield concentrations are evaluated at a time that the $\ce{H2}$ concentration has attained its equilibrium value.  The particular optimization target in Eq.~\eqref{error2} based on methane and hydrogen yields is intended to represent the kinds of limited data that are available in chemical vapor deposition experiments. 

\end{appendix}

\end{document}